\documentclass[3p,times,twocolumn]{elsarticle}
\biboptions{comma,sort&compress}

\usepackage{graphicx}
\usepackage{amsmath}
\usepackage{here}
\usepackage{ecrc}
\volume{00}

\firstpage{1}

\journalname{Nuclear and Particle Physics Proceedings}
\runauth{Stephan Narison}

\jid{nppp}
\jnltitlelogo{Nuclear and Particle Physics Proceedings}
\usepackage{amssymb}
\usepackage[figuresright]{rotating}
\usepackage[colorlinks,linkcolor=blue,anchorcolor=blue,citecolor=blue]{hyperref}
\begin{document}

\begin{frontmatter}

%%
%%%%%%%%%%%%%%%%%%%%%%%%%%%%%%%%%%%%%%%%%%%%%%%%%
%\begin{document}
\title{
%$\la g^3f_{abc} G^aG^bG^c\ra$
%$\la g^3 f_{abc}G^3\ra$ 
% 
%$\alpha_s $, $\la \alpha_sG^2\ra$, $\overline{m}_{c,b}$ and $f_{B_c}$
%from  relativistic heavy quark sum rules$^*$} 
Hadronic Decays of Charmed Hadrons at BESIII\,$^*$} 
 
\cortext[cor0]{Mini-Review talk presented at QCD21, 24th International Conference in QCD (05-09/07/2021, Montpellier - FR). }

\author[label1]{Xinhai Xie\corref{cor1}}
\address[label1]{Peking University, Beijing 100871, People’s Republic of China\\
and\\
State Key Laboratory of Nuclear Physics and Technology, Peking University, Beijing 100871, People’s Republic of China
}
\cortext[cor1]{On behalf of the BESIII collaboration.}
\ead{xiexh\_phys@pku.edu.cn}

\pagestyle{myheadings}
\markright{ }
\begin{abstract}
\noindent
The BESIII experiment has collected the $e^+e^-$ collision data samples corresponding to integrated luminosities of 2.93 $\mathrm{fb}^{-1}$, $3.19$ $\mathrm{fb}^{-1}$, 3.13 $\mathrm{fb}^{-1}$ and 567 $\mathrm{pb}^{-1}$ at center-of-mass energies of 3.773 GeV, 4.178 GeV, 4.189-4.226 GeV and 4.599 GeV, respectively. We report the measurements of strong-phase parameters based on the decays $D^0\to K^0_{S/L}\pi^+\pi^-$, $D^0\to K^0_{S/L}K^+K^-$, $D^0\to K^-\pi^+\pi^+\pi^-$ and $D^0\to K^-\pi^+\pi^0$, which are important input for binned model-independent measurement of the CKM angle $\gamma/\phi_3$. In addition, we report amplitude analyses and branching fraction measurements of $D^+$, $D_s^+$ and $\Lambda_c^+$ decays along with the $\Lambda_c^+$ spin determination.
 
%% keywords
%\ccode{Pac numbers: 11.55.Hx, 12.38.Lg, 13.20-Gd, 14.65.Dw, 14.65.Fy, 14.70.Dj}  
\end{abstract}
\end{frontmatter}
%%%%%%%%%%%%%%%%%%%%%%%%%%%%%%%%%%
%\end{document}
%%%%%%%%%%%%%%%%%%%%%%%%%%%%%%%%%%
%\vspace*{-1.5cm}
\section{Introduction}
%\vspace*{-0.25cm}
 %\nin
%%%%%%%%%%%%%%%%%%%%%%%%%%%%%%%%%%%
The hadronic decays of charmed hadrons play important roles in charm physics and are challenging both theoretically and experimentally, due to non-perturbative effects and great demand of large data sample, respectively.
The BESIII detector records symmetric $e^+e^-$ collisions provided by the BEPCII storage ring, which operates at the center-of-mass energy range from 2.0 to 4.9 $\mathrm{GeV}$. Taking advantage of the advanced performance of the BESIII detector, large data samples of charmed hadron pairs are collected. The charmed meson pairs $D\bar{D}$ ($D_s\bar{D}_s^*$) are produced near the threshold energy of pair production 3.773 GeV~\cite{energy1} (4.178 GeV and 4.189-4.226 GeV~\cite{energy2}), and the charmed baryon pairs $\Lambda_c^+\bar{\Lambda}_c^-$ are produced at pair production energy 4.599 GeV~\cite{energy3}. Based on these large data samples, hadronic decays of charmed hadron can be studied with clean background without accompanying hadrons. The double tag (DT) method is used for most of analyses, where the hadron pair is fully reconstructed. Except for Sec~\ref{sec:Lc}, the single tag (ST) method is used, where only one side of hadron pair is reconstructed. Charge-conjugate modes are implied throughout this paper.

\section{Measurements of strong-phase parameters in $D^0$ decays}
The mechanism of CP violation is of primary importance in particle physics, which is often studied with Cabiboo-Kobayashi-Maskawa (CKM) matrix in the standard model (SM). The measurement of the CKM angle $\gamma/\phi_3$ plays important role in testing the SM describing the CP violation and probing for possible new physics. Currently, the world average value for direct measurements of $\gamma/\phi_3$ have not yet achieved the precision compared with the indirect global fit~\cite{phase1}. As the measurement of the strong-phase parameters of $D^0$ decays can be important input for improving the precision in $\gamma/\phi_3$ measurement in $B^-\to D K^-$ decay, BESIII has reported several measurements for strong-phase parameters.
\subsection{$D^0\to K^0_{S/L}\pi^+\pi^-$ and $D^0\to K^0_{S/L}K^+K^-$}
BESIII has reported the most precise measurements to date of the strong-phase parameters in $D^0\to K^0_{S/L}\pi^+\pi^-$ decays~\cite{phase1}. The GGSZ approach is applied~\cite{GGSZ}, and three different binning schemes are used~\cite{3bin}, which are equal $\Delta\delta_D$ ($\delta_D$ is the relative strong-phase difference between $D$ and $\bar{D}$), optimal and modified optimal binnings. The results of $c_i$ and $s_i$ ($c_i$ and $s_i$ are the average $\cos\delta_D$ and $\sin\delta_D$ in the $i$-th Dalitz bin) for the decay $D\to K^0_{S}\pi^+\pi^-$ is shown in Fig.~\cite{phase1}, which can reduce the associated uncertainty on $\gamma/\phi_3$ measurement from $\sim\!4^\circ$ to $\sim\!1^\circ$ in $B^-\to D(K_S^0\pi\pi)K^-$ in GGSZ approach.
\begin{figure}[tph]
\centering
\includegraphics[trim = 9mm 0mm 0mm 0mm, width=0.48\textwidth]{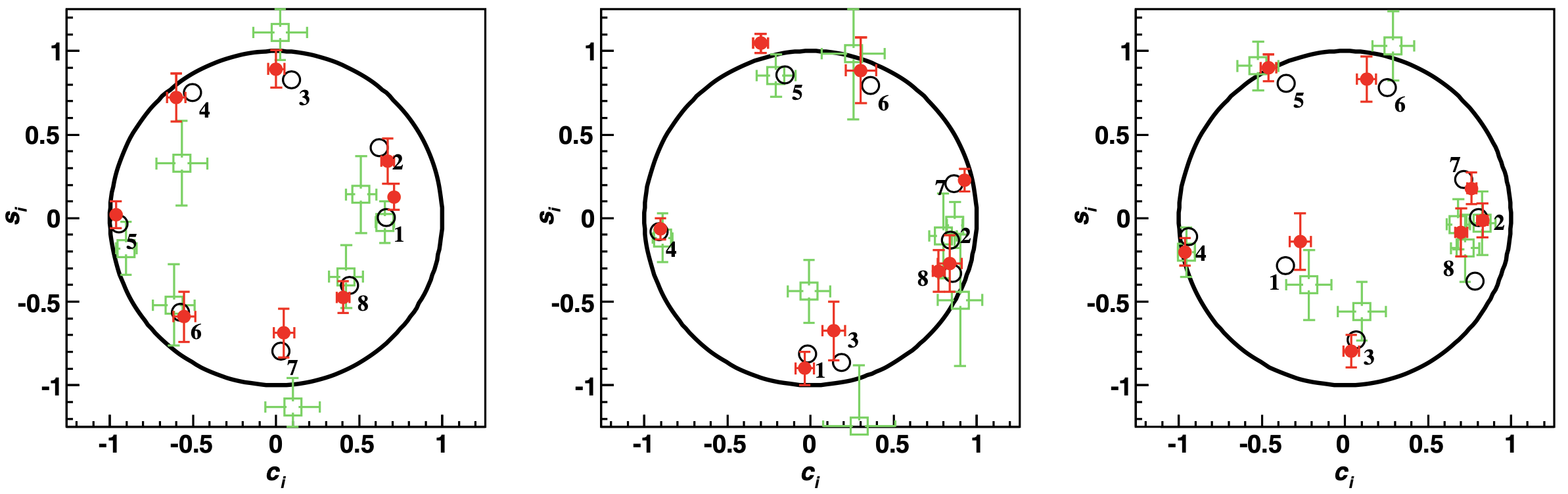}
\caption{Cited from Ref.~\cite{phase1}. The $c_i$ and $s_i$ measured by BESIII (red dots with error bars), the predictions of Ref.\cite{cite1} (black open circles) and the results of Ref.~\cite{3bin}(green open squares with error bars). The left, middle and right plots are from the equal $\Delta\delta_D$, optimal and modified optimal binnings, respectively. The circle indicates the boundary of the physical region $c_i^2+s_i^2=1$.}
\label{fig:phase1}
\end{figure}

\begin{figure}[tph]
\centering
\includegraphics[trim = 9mm 0mm 0mm 0mm, width=0.48\textwidth]{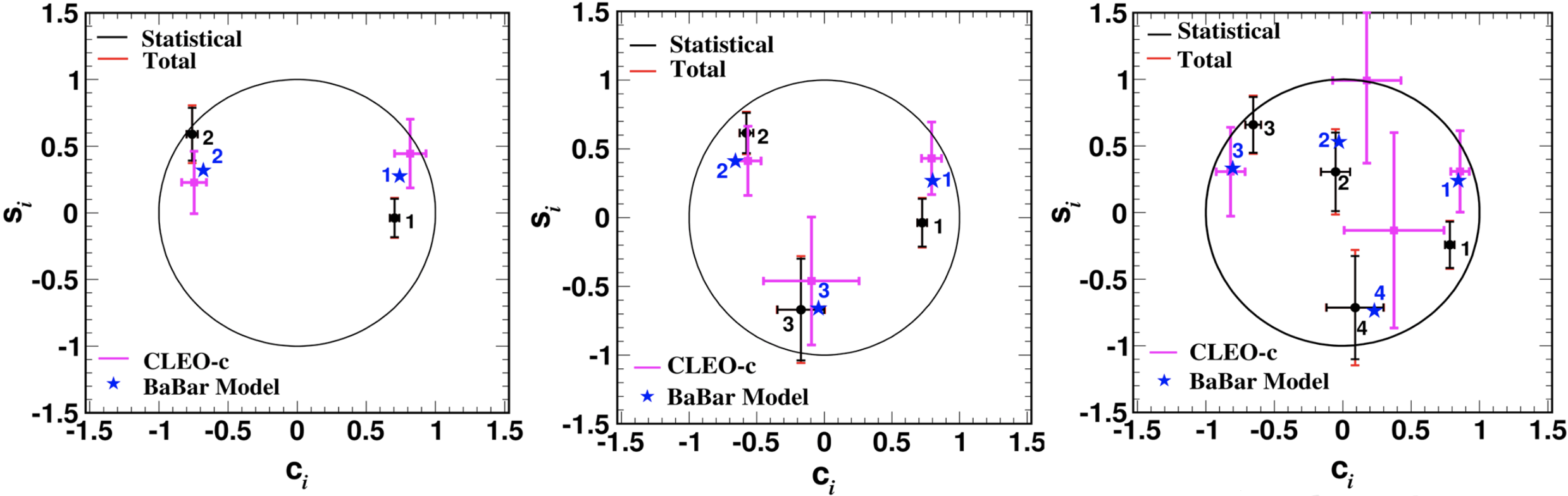}
\caption{Cited from Ref.~\cite{phase2}. The $c_i$ and $s_i$ measured by BESIII (the black points with error bars), the results from CLEO~\cite{3bin} (pink points with error bars) and the predictions of the model reported by the BABAR~\cite{BABAR1} (blue stars). The left, middle and right plots are from the equal $\Delta\delta_D$ binning scheme with $N=2,3$ and 4 bins, respectively. The circle indicates the boundary of the physical region $c_i^2+s_i^2=1$.}
\label{fig:phase2}
\end{figure}

For the decay  $D\to K^0_{S/L}K^+K^-$, BESIII also has reported the most precise measurements to date of the strong-phase parameters~\cite{phase2}. The GGSZ approach is applied~\cite{GGSZ} and the equal $\Delta\delta_D$ binning scheme with $N=2,3,4$ bins are used. The results of $c_i$ and $s_i$ are shown in Fig.~\ref{fig:phase2}, which result in the associated uncertainty on $\gamma/\phi_3$ measurement as $\sim\!2.3^\circ$, $\sim\!1.3^\circ$ and $\sim\!1.3^\circ$ for bins $N=2,3$ and 4

\subsection{$D\to K^-\pi^+\pi^+\pi^-$ and $D\to K^-\pi^+\pi^0$}
\begin{figure}[tph]
\centering
\includegraphics[trim = 9mm 0mm 0mm 0mm, width=0.48\textwidth]{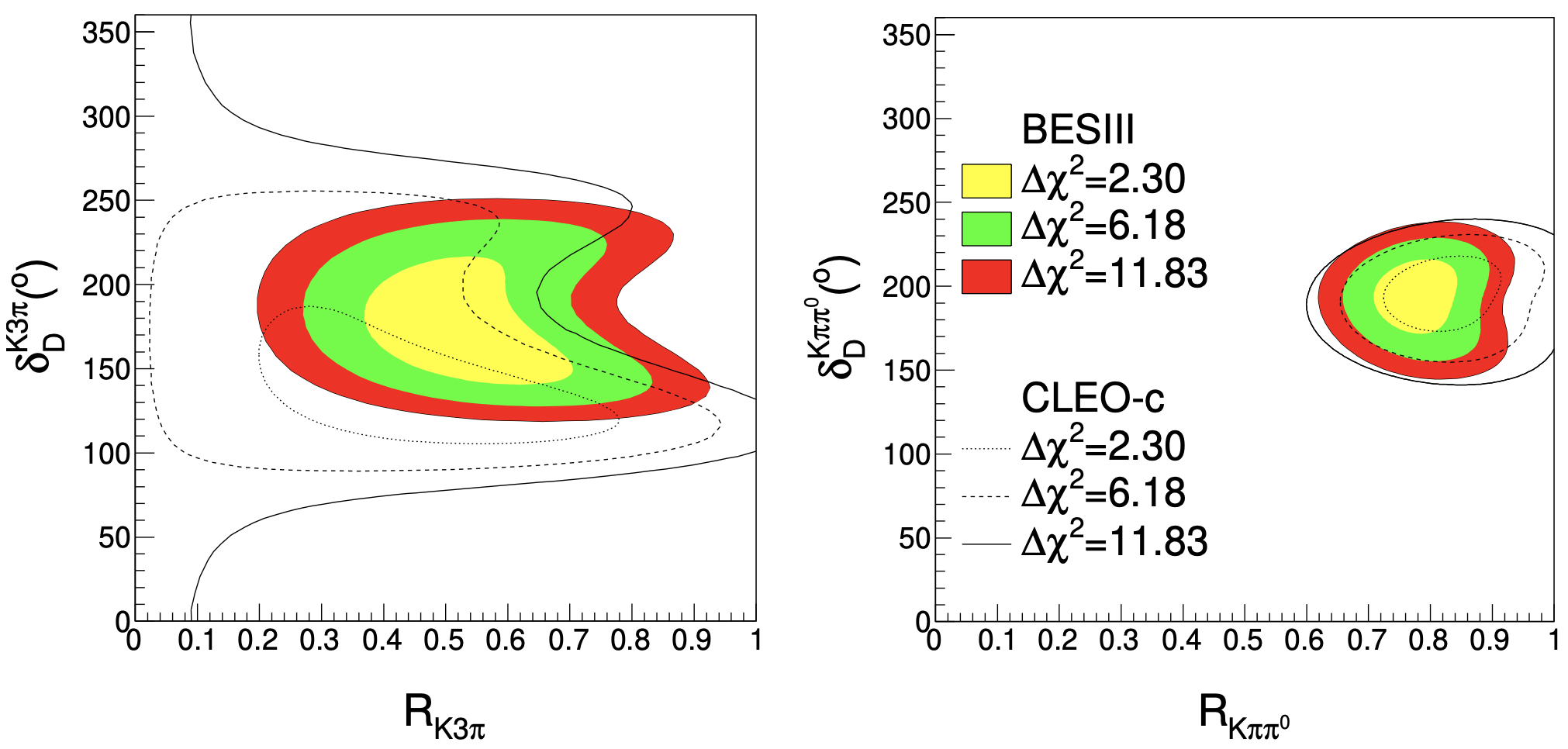}
\caption{Cited from Ref.~\cite{phase3}. Scans of $\Delta\chi^2$ in the global $(R_{K3\pi},\delta_D^{K3\pi})$ and $R_{K\pi\pi^0},\delta_D^{K\pi^0}$ parameter space, showing the $\Delta\chi^2=2.30, 6.18, 11.83$ intervals, which correspond to 68.3\%, 95.4\% and 99.7\% confidence levels in the two-dimensional parameter space. Also shown are the equivalent contours determined from the CLEO-c data~\cite{cite2}.}
\label{fig:phase3}
\end{figure}

Recently, BESIII has reported improved measurements of the coherence factors and average strong-phase differences in the decays $D\to K^-\pi^+\pi^+\pi^-$ and $D\to K^-\pi^+\pi^0$~\cite{phase3}. Using global analysis and equal $\Delta\delta_D$ binning scheme with $N=4$ bins~\cite{4bin}, the coherence factors are determined to be $R_{K3\pi}=0.52^{+0.12}_{-0.10\,\mathrm{tot.}}$ and $R_{K\pi\pi^0}=0.78\pm 0.04_\mathrm{tot.}$, with values for the average strong-phase differences to be $\delta_D^{K3\pi}=\left(167^{+31}_{-19\,\mathrm{tot.}}\right)^\circ$ and $\delta_D^{K\pi^0}=\left(196^{+14}_{-15\,\mathrm{tot.}}\right)^\circ$. The scan plots of $\Delta\chi^2$ in the global $(R_{K3\pi},\delta_D^{K3\pi})$ and $R_{K\pi\pi^0},\delta_D^{K\pi^0}$ parameter space are shown in Fig.~\ref{fig:phase3}, and the region of parameter space $(R_{K3\pi},\delta_D^{K3\pi})$ encompassed by the 2$\sigma$ and 3$\sigma$ confidence intervals is significantly more constrained.

\section{Measurements in $D^+$ decay rates}
Doubly Cabibbo-suppressed (DCS) decays of $D$ mesons can provide unique test for the weak decay mechanisms in charmed sector. BESIII has reported two independent measurements for the DCS decay $D^+ \to K^+\pi^+\pi^-\pi^0$ using different $D^-$ tag modes.

Using three Cabbibo-favored (CF) tag modes $D^-\to K^+\pi^-\pi^-,K_S^0\pi^-$ and $K^+\pi^-\pi^-\pi^0$~\cite{DCS1}, the branching fraction (BF) is measured to be $\mathcal{B}_{D^+\to K^+\pi^+\pi^-\pi^0}=(1.13\pm 0.08_\mathrm{stat.}\pm0.03_\mathrm{syst.}) \times 10^{-3}$, where the decays containing narrow intermediate resonances including $D^+\to K^+\eta,K^+\omega$ and $K^+\phi$ have been removed. The relative ratio is obtained to be $\mathcal{B}_{D^+\to K^+\pi^+\pi^-\pi^0}/\mathcal{B}_{D^+\to K^-\pi^+\pi^+\pi^0}=(1.81\pm0.15_\mathrm{tot.})\%$, which corresponds to be $(6.28\pm0.52_\mathrm{tot.})\tan^4\theta_C$, where $\theta_C$ is the Cabibbo mixing angle. This ratio is significantly larger than the values found in most other DCS decays around (0.21$\sim$0.58)\%, which sheds indicate of massive isospin symmetry violation in the decays $D^+ \to K^+\pi^+\pi^-\pi^0$ and $D^0 \to K^+\pi^-\pi^-\pi^+$. Besides, no CP violation evidence is observed in this DCS decay.

Using the semi-leptonic tag modes $D^-\to K^0 e^- \bar{\nu}_e$ and $K^+\pi^-e^- \bar{\nu}_e$~\cite{DCS2}, the independent measurement is performed. The BF is measured to be $\mathcal{B}_{D^+\to K^+\pi^+\pi^-\pi^0}=(1.03\pm 0.12_\mathrm{stat.}\pm0.06_\mathrm{syst.}) \times 10 ^{-3}$ subtracting $\eta,\omega$ and $\phi$ contributions. The relative ratio can also obtained to be $\mathcal{B}_{D^+\to K^+\pi^+\pi^-\pi^0}/\mathcal{B}_{D^+\to K^-\pi^+\pi^+\pi^0}=(1.65\pm0.21_\mathrm{tot.})\%$ corresponding to $(5.73\pm0.73_\mathrm{tot.})\tan^4\theta_C$, which are consistent and confirmed the results using CF tag modes.

\section{Amplitude analyses in $D^+$ and $D_s^+$ decays}
Based on the large $D^+D^-$ and $D_s\bar{D}_s^*$ pair data samples, BESIII has reported amplitude analyses for several $D^+$ and $D_s^+$ meson multi-body decays.

\subsection{Amplitude analysis of $D^+ \to K^+K_S^0\pi^0$}
For the singly Cabibbo-suppressed (SCS) decay $D^+ \to K^+K_S^0\pi^0$, BESIII has reported the first amplitude analysis results of this decay~\cite{kskpi0} using the data sample of 692 signal candidates with 97.4\% purity. The process $D^+\to K^*(892)^+K_S^0$ is found to be dominant with a fit fraction (FF) of $(57.1\pm2.6_\mathrm{stat.}\pm4.2_\mathrm{syst.})\%$, and the absolute BF is measured to be $\mathcal{B}_{D^+\to K^*(892)^+K_S^0}=(8.69\pm0.40_\mathrm{stat.}\pm0.64_\mathrm{syst.}\pm0.51_\mathrm{Br.})\times 10^{-3}$, whose precision is significantly improved. Besides, the isospin symmetric process $D^+\to \bar{K}^*(892)^0 K^+$ is found to be with a FF of $(10.2\pm1.5_\mathrm{stat.}\pm2.2_\mathrm{syst.})\%$ and the absolute BF is measured to be $\mathcal{B}_{D^+\to \bar{K}^*(892)^0 K^+}=(3.10\pm0.46_\mathrm{stat.}\pm0.68_\mathrm{syst.}\pm0.18_\mathrm{Br.})\times 10^{-3}$, which is consistent with the previous result~\cite{PDG}.

\subsection{$D_s^+\to K_S^0K^-\pi^+\pi^+$}
For the decay $D_s^+\to K_S^0K^-\pi^+\pi^+$, the first amplitude analysis for this mode is reported using the data sample of 1318 signal candidates with around 94\% purity~\cite{Ds1}. The resonant process $D_s^+\to K^*(892)^+\bar{K}^*(892)^0$ is found to be dominant with a FF of $(40.6\pm2.9_\mathrm{stat.}\pm4.9_\mathrm{syst.})\%$ and the corresponding absolute BF is obtained to be $(5.34\pm0.39_\mathrm{stat.}\pm0.64_\mathrm{syst.})\%$. Using the results from the amplitude analysis, the BF is measured to be $(1.46\pm0.05_\mathrm{stat.}\pm0.05_\mathrm{syst.})\%$ with improved precision.

\subsection{$D_s^+\to K_S^0\pi^+\pi^0$}
For the decay $D_s^+\to K_S^0\pi^+\pi^0$, the first amplitude analysis is performed based on the data sample of 609 signal candidates with 83.1\% purity~\cite{Ds2}. The dominant resonant process is found to be $D_s^+\to K_S^0\rho^+$, with a FF of $(50.2\pm7.2_\mathrm{stat.}\pm3.9_\mathrm{syst.})\%$, and the corresponding absolute BF is obtained to be $\mathcal{B}_{D_s^+\to K_S^0\rho^+}=(2.73\pm0.42_\mathrm{stat.}\pm0.22_\mathrm{syst.})\times 10^{-3}$. For the other isospin symmetric processes $D_s^+\to K^*(892)\pi$, the absolute BFs are also measured to be $\mathcal{B}_{D^+\to K^*(892)^0\pi^+}=(0.45\pm0.12_\mathrm{stat.}\pm0.05_\mathrm{syst.})\times 10^{-3}$ and $\mathcal{B}_{D^+\to K^*(892)^+\pi^0}=(0.21\pm0.09_\mathrm{stat.}\pm0.03_\mathrm{syst.})\times 10^{-3}$. Using the model from amplitude analysis, the absolute BF for the decay $D_s^+\to K_S^0\pi^+\pi^0$ is measured to be $(5.43\pm0.30_\mathrm{stat.}\pm0.15_\mathrm{syst.})\times 10^{-3}$ with improved precision.

\subsection{$D_s^+\to \eta\pi^+\pi^+\pi^-$}
For the decay $D_s^+\to \eta\pi^+\pi^+\pi^-$, the first amplitude analysis is reported using the data sample of 1306 signal candidates with purity up to 85\%~\cite{Ds3}. The dominant resonant process is the cascade process $D_s^+\to a_1(1260)^+,a_1(1260)^+\to\rho^0\pi^+$ with a FF of $(55.4\pm3.9_\mathrm{stat.}\pm2.0_\mathrm{syst.})\%$. The corresponding absolute BF is obtained to be $\mathcal{B}_{D_s^+\to a_1(1260)^+,a_1(1260)^+\to\rho^0\pi^+}=(1.73\pm0.14_\mathrm{stat.}\pm0.08_\mathrm{syst.})\%$. For another important W-annihilation process $D_s^+\to a_0(980)^+\rho^0,a_0(980)^+\to\eta\pi^+$ which plays a key role in understanding the nature of $a_0(980)^+$, it is also found to have significant contribution in the amplitude analysis, and the absolute BF is obtained to be $\mathcal{B}_{D_s^+\to a_0(980)^+\rho^0,a_0(980)^+\to\eta\pi^+}=(0.21\pm0.08_\mathrm{stat.}\pm0.05_\mathrm{syst.})\%$, which is larger than the BFs of other measured pure $W$-annihilation decays by one order of magnitude. Besides, using the results from amplitude analysis, the BF for the decay $D_s^+\to \eta\pi^+\pi^+\pi^-$ is also measured to be $(3.12\pm0.13_\mathrm{stat.}\pm0.09_\mathrm{syst.})\%$.

\subsection{$D_s^+\to K^+K^-\pi^+$}
For the decay $D_s^+\to K^+K^-\pi^+$, an amplitude analysis is performed based on the data sample of 4399 signal candidates with 99.6\% purity~\cite{Ds4}. Due to large overlap of $a_0(980)\to K^+K^-$ and $f_0(980)\to K^+K^-$ sharing the same spin-parity, in the amplitude analysis, the $K^+K^-$ low-mass resonances are not distinguished and considered as a combined state parametrized using a model-independent partial wave analysis (MIPWA). With the detection efficiency based on the amplitude analysis results, the absolute BF is measured to be $\mathcal{B}_{D_s^+\to K^+K^-\pi^+}=(5.47\pm0.08_\mathrm{stat.}\pm0.13_\mathrm{syst.})\%$ which is currently the most precise measurement.

\subsection{$D_s^+\to K^+K^-\pi^+\pi^0$}
For the decay $D_s^+\to K^+K^-\pi^+\pi^0$, the first amplitude analysis is reported using the data sample of 3088 signal candidates with 97.5\% purity~\cite{Ds5}. Two $D_s^+\to VV$ modes are found to be dominant, where $V$ denotes vector meson, which are $D_s^+\to \phi\rho^+$ and $K^*(892)^+\bar{K}^*(892)^0$, and their BFs are measured to be $\mathcal{B}_{D_s^+\to \phi\rho^+}=(5.59\pm0.15_\mathrm{stat.}\pm0.30_\mathrm{syst.})\%$ and $\mathcal{B}_{D_s^+\to K^*(892)^+\bar{K}^*(892)^0}=(5.64\pm0.23_\mathrm{stat.}\pm0.27_\mathrm{syst.})\%$, whose precision is much improved. Knowledge of another interesting resonance, $K_1(1270)^0$, can also be studied in this amplitude analysis through the process $D_s^+\to \bar{K}_1(1270)^0K^+$ where $\bar{K}_1(1270)^0$ decays into $K^-\rho^+$ or $K^*(892)\pi$. The relative ratio $R_{K_1(1270)^0}\equiv\mathcal{B}_{K_1(1270)^0\to K^*(892)\pi}/\mathcal{B}_{K_1(1270)^0\to K\rho}$ is measured to be $0.99\pm0.15_\mathrm{stat.}\pm0.18_\mathrm{syst.}$, which is a valuable crosscheck for other experiments. Besides, using the detection efficiency based on the amplitude analysis results, the absolute BF is measured to be $\mathcal{B}_{D_s^+\to K^+K^-\pi^+\pi^0}=(5.42\pm0.10_\mathrm{stat.}\pm0.17_\mathrm{syst.})\%$, whose precision is much improved.

\subsection{$D_s^+\to \pi^+\pi^+\pi^-$}
\begin{figure}[tph]
\centering
\includegraphics[trim = 9mm 0mm 0mm 0mm, width=0.48\textwidth]{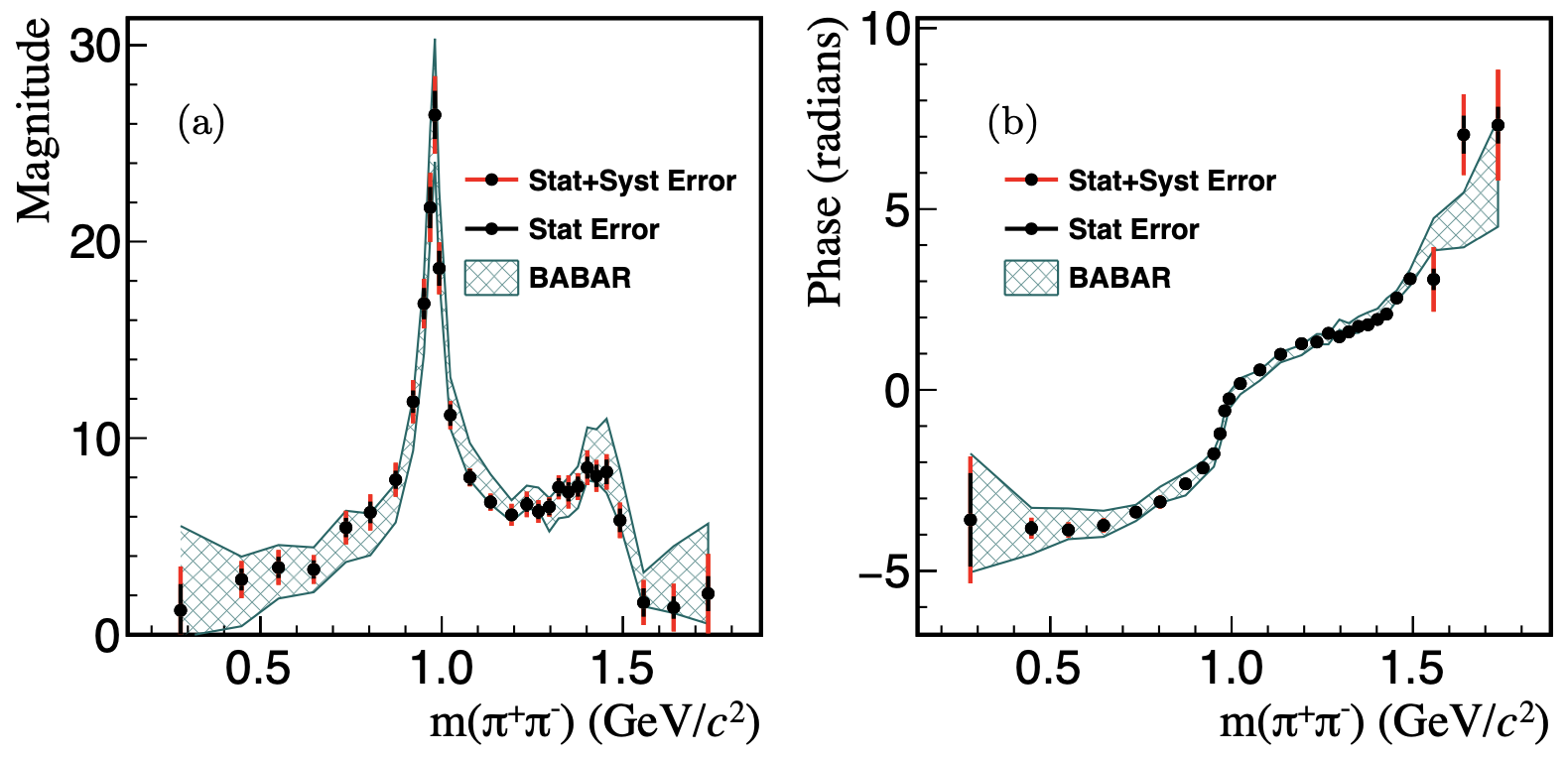}
\caption{Cited from Ref.~\cite{Ds6}. (a) Magnitudes and (b) phases of the $\mathcal{S}$-wave control points. The results are compared to the BABAR results~\cite{babar} with the same binning scheme and similar data sample size.}
\label{fig:swave}
\end{figure}

For the decay $D_s^+\to \pi^+\pi^+\pi^-$, an amplitude analysis is reported based on the data sample of 13797 signal candidates with around 80\% purity~\cite{Ds6}. Due to possible overlap of the complicated $\pi^+\pi^-$ components, the MIPWA is used to parametrized the $\pi^+\pi^-$ $\mathcal{S}$-wave components, while the $\mathcal{P}$-wave and $\mathcal{D}$-wave components are described with the normal relativistic Breit-Wigner amplitudes. In this amplitude analysis, the $\pi^+\pi^-$ $\mathcal{S}$-wave is dominant with a FF of $(84.2\pm0.8_\mathrm{stat.}\pm1.3_\mathrm{syst.})\%$, where the $\mathcal{S}$-wave control points are shown in Fig.~\ref{fig:swave}, which agree with previous BABAR results~\cite{babar} and with higher precision.

\section{Studies on $\Lambda_c^+$ decays}
\label{sec:Lc}
\subsection{Branching fraction of $\Lambda_c^+\to p K_S^0 \eta$}
\begin{figure}[tph]
\centering
\includegraphics[trim = 9mm 0mm 0mm 0mm, width=0.48\textwidth]{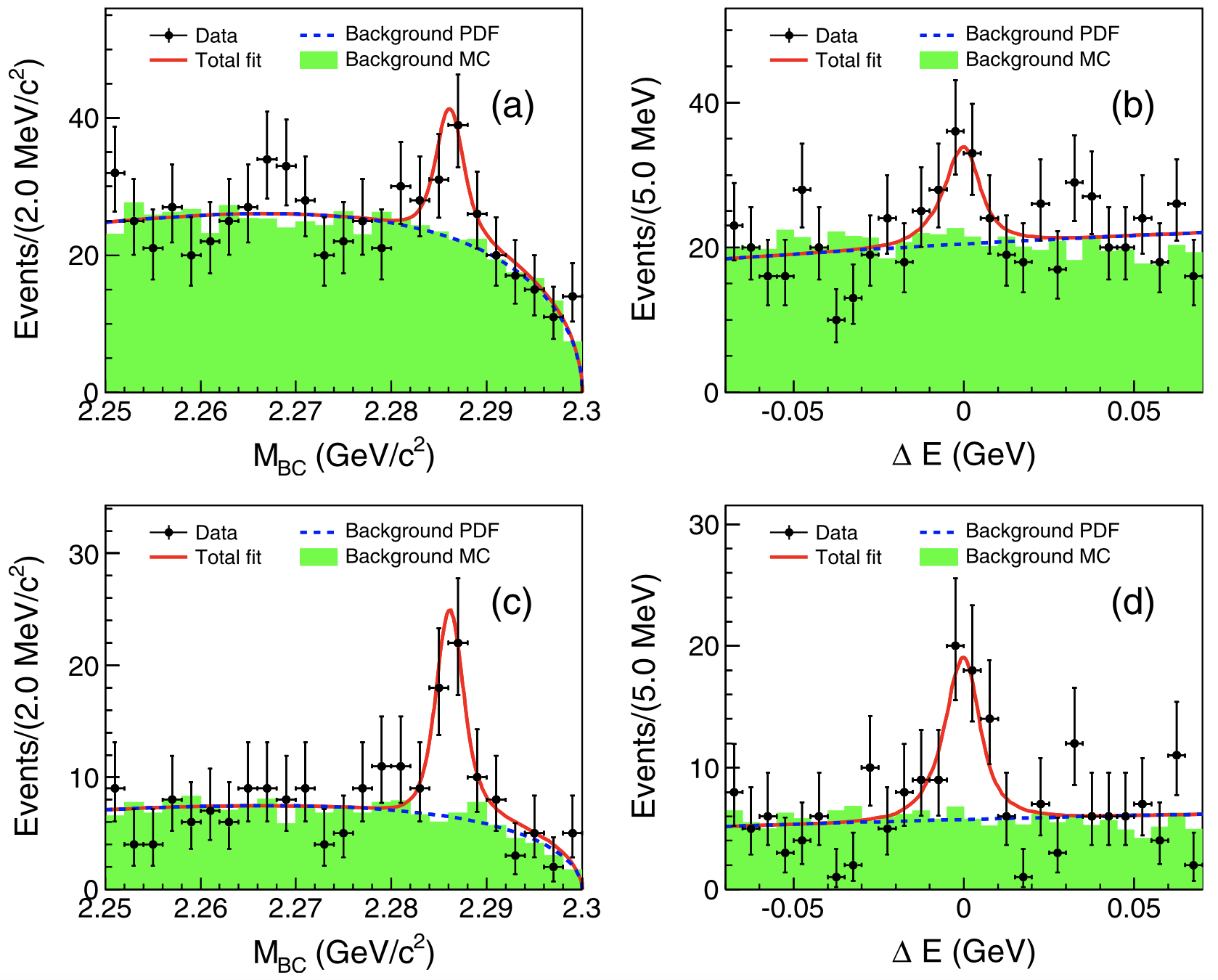}
\caption{Cited from Ref.~\cite{Lc1}. One-dimensional projection plots of (a) $M_\mathrm{BC}$ and (b) $\Delta E$ distributions, as well as background reduced projection plots of (c) $M_\mathrm{BC}$ ($|\Delta E|<0.02\,\mathrm{GeV}$) and (d) $\Delta E$ ($2.28 <M_\mathrm{BC} <2.295\,\mathrm{GeV/}c^2$) with the fit results. Points with error bars are data, the solid red line is the total fit result, the blue-dashed line is the background contribution, and the green histogram is the background distribution from the inclusive MC sample.}
\label{fig:pKseta}
\end{figure}

Recently, BESIII has reported the first BF measurement of the decay $\Lambda_c^+\to p K_S^0 \eta$ using the ST method~\cite{Lc1} .The signal yield is extracted by performing two-dimensional fit on the beam-constrained mass $M_\mathrm{BC}$, and the energy difference $\Delta E$, which can be found in Fig.~\ref{fig:pKseta}. The statistical significance for the signal is found to be 5.3$\sigma$ and the absolute BF is measured to be $\mathcal{B}_{\Lambda_c^+\to p K_S^0 \eta}=(0.414\pm0.084_\mathrm{stat.}\pm0.028_\mathrm{syst.})\%$ for the first time.

\subsection{$\Lambda_c^+$ spin determination}
\begin{figure}[tph]
\centering
\includegraphics[trim = 9mm 0mm 0mm 0mm, width=0.48\textwidth]{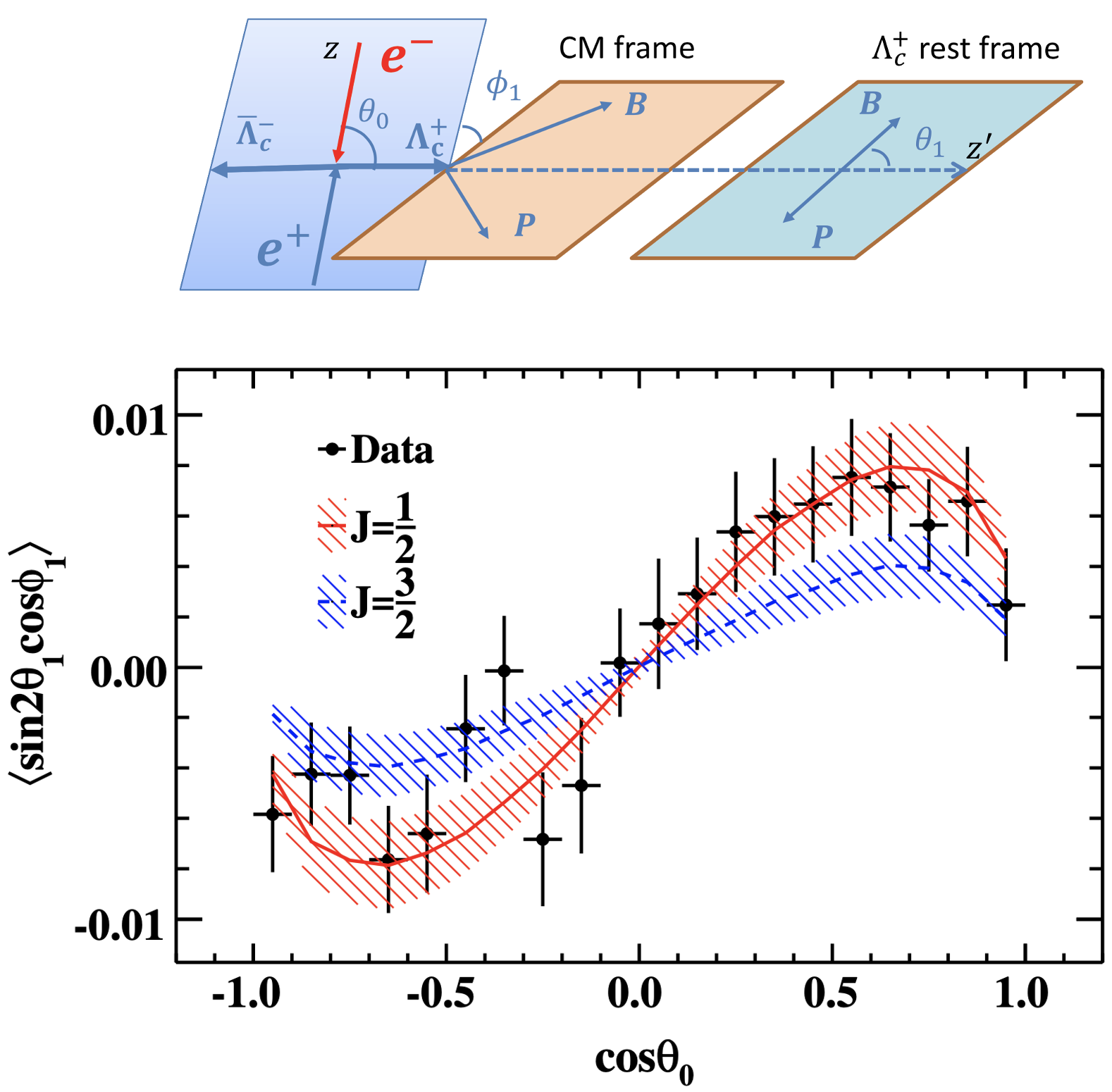}
\caption{Cited from Ref.~\cite{LcSpin}. The upper subfigure shows the definition of the helicity frame for $e^+e^-\to\Lambda_c^+\bar{\Lambda}_c^-,\Lambda_c^+\to BP$, where $B$ and $P$ denote a spin $J=1/2$ baryon and a meson, respectively. The lower subfigure shows the moments $\langle\sin2\theta_1\cos\phi_1\rangle$ as a function of $\cos\theta_0$.}
\label{fig:LcSpin}
\end{figure}

Since the $\Lambda_c^+$ baryon has been firstly discovered for more than 30 years, no spin determination of the $\Lambda_c^+$ is performed yet, and the $\Lambda_c^+$ is inferred to have spin $1/2$ from the na\"ive quark model. In order to test this spin assignment in experiment and to test the quark model hadron classification, BESIII recently reported the results of $\Lambda_c^+$ spin determination~\cite{LcSpin}. The Multi-dimensional angular analysis on the ST samples of $\Lambda_c^+\to pK_S^0$ are carried out to test both hypotheses of $J= 1/2$ and $3/2$. The definition of the helicity frame and the angular distribution are shown in Fig.~\ref{fig:LcSpin}, which give the conclusion that $J= 1/2$ is preferred over $J= 3/2$ with significance of around 6$\sigma$ and is consistent with the expectation of na\"ive quark model. This will be a cornerstone in the extraction of the properties of heavier charmed and beauty baryons.

\section{Summary}
Based on the data samples collected by the BESIII collaboration, we reported the strong-phase parameters in $D^0$ decays with the highest precision to date, which can reduce the systematic uncertainty for CKM angle $\gamma/\phi_3$ measurements at LHCb and Belle II. For $D^+$ decays, we reported two independent measurements for the DCS decay $D^+ \to K^+\pi^+\pi^-\pi^0$ and the amplitude analysis for the decay $D^+\to K^+K_S^0\pi^0$. For $D_s^+$ decays, we reported amplitude analyses for six decay modes and the corresponding measurement for BFs. For $\Lambda_c^+$ decays, we reported the first measurement of the decay $\Lambda_c^+\to p K_S^0 \eta$ and the spin determination of the $\Lambda_c^+$ baryon. All above results can help us deep understand the CP violation effects and internal dynamics in charm hadron decays~\cite{Li:2021iwf,Saur:2020rgd}. In the future, BESIII plans to take another 17 $\mathrm{fb}^{-1}$ at center-of-mass energies of $3.773\,\mathrm{GeV}$~\cite{white} and more results in hadronic charm meson decays can be expected.

\vfill\eject
%%%%%%%%%%%%%%%

\end{document}